\documentclass[sigconf]{acmart}
\AtBeginDocument{%
  \providecommand\BibTeX{{%
    \normalfont B\kern-0.5em{\scshape i\kern-0.25em b}\kern-0.8em\TeX}}}

\setcopyright{iw3c2w3}
\copyrightyear{2020}
\acmYear{2020}
\acmDOI{10.1145/3366424.3383525}

\acmConference[WWW '20]{Proceedings of The Web Conference 2020}{April 20--24, 2020}{Taipei, Taiwan}
\acmBooktitle{Proceedings of The Web Conference 2020 (WWW '20), April 20--24, 2020, Taipei, Taiwan} 
\acmPrice{}
\acmISBN{978-1-4503-7023-3/20/04}

\begin{document}

\title{QnAMaker: Data to Bot in 2 Minutes}




\author{Parag Agrawal}
\affiliation{\institution{Microsoft Corporation}}
\email{paragag@microsoft.com}

\author{Tulasi Menon}
\affiliation{\institution{Microsoft Corporation}}
\email{tulasim@microsoft.com}

\author{Aya Kamel}
\affiliation{\institution{Microsoft Corporation}}
\email{aykame@microsoft.com}

\author{Michel Naim}
\affiliation{\institution{Microsoft Corporation}}
\email{mgerguis@microsoft.com}

\author{Chaikesh Chouragade}
\affiliation{\institution{Microsoft Corporation}}
\email{chchoura@microsoft.com}

\author{Gurvinder Singh}
\affiliation{\institution{Microsoft Corporation}}
\email{gurvsing@microsoft.com}

\author{Rohan Kulkarni}
\affiliation{\institution{Microsoft Corporation}}
\email{rokulka@microsoft.com}

\author{Anshuman Suri}
\affiliation{\institution{Microsoft Corporation}}
\email{ansuri@microsoft.com}

\author{Sahithi Katakam}
\affiliation{\institution{Microsoft Corporation}}
\email{sakataka@microsoft.com}

\author{Vineet Pratik}
\affiliation{\institution{Microsoft Corporation}}
\email{vipratik@microsoft.com}

\author{Prakul Bansal}
\affiliation{\institution{Microsoft Corporation}}
\email{prabansa@microsoft.com}

\author{Simerpreet Kaur}
\affiliation{\institution{Microsoft Corporation}}
\email{sikaur@microsoft.com}

\author{Neha Rajput}
\affiliation{\institution{Microsoft Corporation}}
\email{nerajput@microsoft.com}

\author{Anand Duggal}
\affiliation{\institution{Microsoft Corporation}}
\email{anduggal@microsoft.com}

\author{Achraf Chalabi}
\affiliation{\institution{Microsoft Corporation}}
\email{achalabi@microsoft.com}

\author{Prashant Choudhari}
\affiliation{\institution{Microsoft Corporation}}
\email{pchoudh@microsoft.com}

\author{Somi Reddy Satti}
\affiliation{\institution{Microsoft Corporation}}
\email{sosatti@microsoft.com}

\author{Niranjan Nayak}
\affiliation{\institution{Microsoft Corporation}}
\email{niranjan@microsoft.com}

\renewcommand{\shortauthors}{Trovato and Tobin, et al.}

\begin{abstract}
Having a bot for seamless conversations is a much-desired feature that products and services today seek for their websites and mobile apps. These bots help reduce traffic received by human support significantly by handling frequent and directly answerable known questions. Many such services have huge reference documents such as FAQ pages, which makes it hard for users to browse through this data. A conversation layer over such raw data can lower traffic to human support by a great margin. We demonstrate QnAMaker, a service that creates a conversational layer over semi-structured data such as FAQ pages, product manuals, and support documents. QnAMaker is the popular choice for Extraction and Question-Answering as a service and is used by over 15,000 bots in production. It is also used by search interfaces and not just bots.

\end{abstract}




\copyrightyear{2020}
\acmYear{2020}
\acmConference[WWW '20 Companion]{Companion Proceedings of the Web Conference 2020}{April 20--24, 2020}{Taipei, Taiwan}
\acmBooktitle{Companion Proceedings of the Web Conference 2020 (WWW '20 Companion), April 20--24, 2020, Taipei, Taiwan}
\acmPrice{}
\acmDOI{10.1145/3366424.3383525}
\acmISBN{978-1-4503-7024-0/20/04}

\maketitle

\section{Introduction}


QnAMaker aims to simplify the process of bot creation by extracting Question-Answer (QA) pairs from data given by users into a Knowledge Base (KB) and providing a conversational layer over it. KB here refers to one instance of azure search index, where the extracted QA are stored.  Whenever a developer creates a KB using QnAMaker, they automatically get all NLP capabilities required to answer user's queries. There are other systems such as Google's Dialogflow, IBM's Watson Discovery which tries to solve this problem. QnAMaker provides unique features for the ease of development such as the ability to add a persona-based chit-chat layer on top of the bot. Additionally, bot developers get automatic feedback from the system based on end-user traffic and interaction which helps them in enriching the KB; we call this feature active-learning\footnote{https://docs.microsoft.com/en-us/azure/cognitive-services/qnamaker/how-to/improve-knowledge-base}. Our system also allows user to add Multi-Turn structure to KB using hierarchical extraction and contextual ranking. QnAMaker today supports over 35 languages, and is the only system among its competitors to follow a Server-Client architecture; all the KB data rests only in the client's subscription, giving users total control over their data.
QnAMaker is part of Microsoft Cognitive Service and currently runs using the Microsoft Azure Stack\footnote{https://azure.microsoft.com/en-us/services/cognitive-services/qna-maker/}.

\section{System description}
\subsection{Architecture}
\begin{figure}
    \centering
    \includegraphics[width=0.4\textwidth]{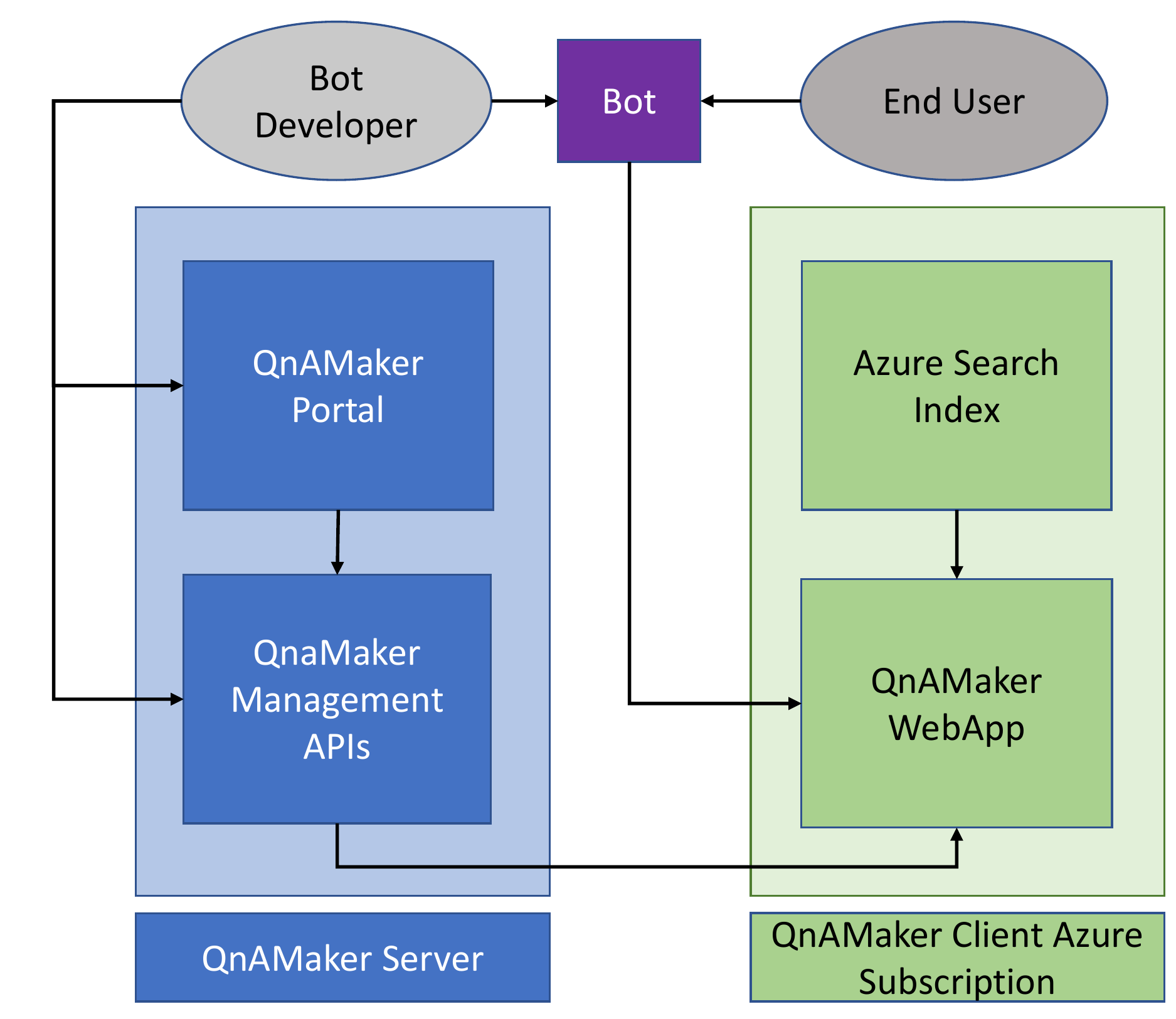}
    \caption{Interactions between various components of QnaMaker, along with their scopes: server-side and client-side}
    \label{fig:architecture}
\end{figure}

As shown in Figure~\ref{fig:architecture}, humans can have two different kinds of roles in the system: \textbf{Bot-Developers} who want to create a bot using the data they have, and \textbf{End-Users} who will chat with the bot(s) created by bot-developers. The components involved in the process are:

\begin{itemize}
\item \textit{QnAMaker Portal}\footnote{https://www.qnamaker.ai/}: This is the Graphical User Interface (GUI) for using QnAMaker. This website is designed to ease the use of management APIs. It also provides a test pane.
\item \textit{QnaMaker Management APIs}: This is used for the extraction of Question-Answer (QA) pairs from semi-structured content. It then passes these QA pairs to the web app to create the Knowledge Base Index. 
\item \textit{Azure Search Index}: Stores the KB with questions and answers as indexable columns, thus acting as a retrieval layer.
\item \textit{QnaMaker WebApp}: Acts as a layer between the Bot, Management APIs, and Azure Search Index. WebApp does ranking on top of retrieved results. WebApp also handles feedback management for active learning.
\item \textit{Bot}: Calls the WebApp with the User's query to get results.
\end{itemize}

\subsection{Bot Development Process}
Creating a bot is a 3-step process for a bot developer:
\begin{enumerate}
    \item Create a QnaMaker Resource in Azure: This creates a WebApp with binaries required to run QnAMaker. It also creates an Azure Search Service for populating the index with any given knowledge base, extracted from user data
    \item Use Management APIs to Create/Update/Delete your KB: The Create API automatically extracts the QA pairs and sends the Content to WebApp, which indexes it in Azure Search Index. Developers can also add persona-based chat content and synonyms while creating and updating their KBs. 
    \item Bot Creation: Create a bot using any framework and call the WebApp hosted in Azure to get your queries answered. There are Bot-Framework templates \footnote{https://docs.microsoft.com/en-us/azure/bot-service/bot-builder-howto-qna} provided for the same.
\end{enumerate}


\subsection{Extraction} \label{sec:extraction}
The Extraction component is responsible for understanding a given document and extracting potential QA pairs. These QA pairs are in turn used to create a KB to be consumed later on by the \textit{QnAMaker WebApp} to answer user queries. First, the basic blocks from given documents such as text, lines are extracted. Then the layout of the document such as columns, tables, lists, paragraphs, etc is extracted. This is done using Recursive X-Y cut~\cite{ha1995recursive}. Following Layout Understanding, each element is tagged as headers, footers, table of content, index, watermark, table, image, table caption, image caption, heading, heading level, and answers. Agglomerative clustering~\cite{cutting1992scatter} is used to identify heading and hierarchy to form an intent tree. Leaf nodes from the hierarchy are considered as QA pairs. In the end, the intent tree is further augmented with entities using CRF-based sequence labeling. Intents that are repeated in and across documents are further augmented with their parent intent, adding more context to resolve potential ambiguity.

\subsection{Retrieval And Ranking}

\begin{figure}
    \centering
    \includegraphics[width=0.4\textwidth]{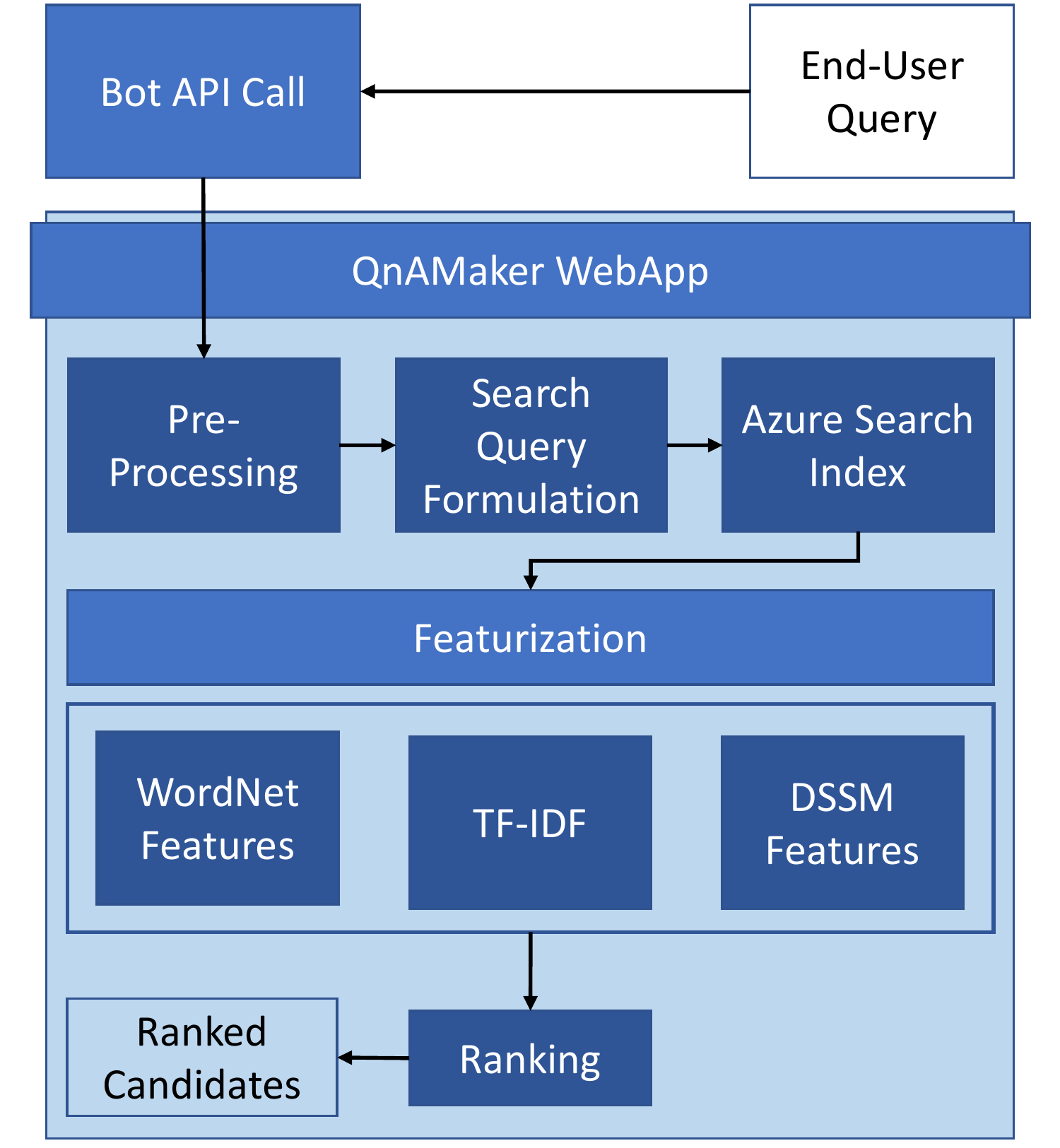}
    \caption{QnAMaker Runtime Pipeline}
    \label{fig:QnaRuntimeFigure}
\end{figure}

QnAMaker uses \textit{Azure Search Index} as it's retrieval layer, followed by re-ranking on top of retrieved results (Figure~\ref{fig:QnaRuntimeFigure}). Azure Search is based on inverted indexing and TF-IDF scores. Azure Search provides fuzzy matching based on edit-distance, thus making retrieval robust to spelling mistakes. It also incorporates lemmatization and normalization. These indexes can scale up to millions of documents, lowering the burden on \textit{QnAMaker WebApp} which gets less than 100 results to re-rank.

Different customers may use QnAMaker for different scenarios such as banking task completion, answering FAQs on company policies, or fun and engagement. The number of QAs, length of questions and answers, number of alternate questions per QA can vary significantly across different types of content. Thus, the ranker model needs to use features that are generic enough to be relevant across all use cases.

\subsubsection{Pre-Processing}
The pre-processing layer uses components such as Language Detection, Lemmatization, Speller, and Word Breaker to normalize user queries. It also removes junk characters and stop-words from the user's query.

\begin{table*}
    \centering
    \begin{tabular}{|c|c|c|c|c|}
    \hline
        Domain & Number of QAs & Avg Questions per QA & AUC (\%) & $F_{1}$ \\ 
        & & & & (top answer) \\
        \hline \hline
        Navigation Help Bot & 56 & 12.5 & 88.7 & 71.2 \\
        \hline
        Chit-Chat Alone & 100 & 9.8 & 92.4 & 88.6 \\
        \hline
        CustomerCare Interface & 164 & 2.2 & 90.9 & 86.7 \\
        \hline
        HR Internal Bot & 52 & 1.0 & 85.5 & 82.6 \\
        \hline
        HR Internal Bot (with Chit-Chat)  & 152 & 6.78 & 82.7 & 77.6 \\
        \hline 
    \end{tabular}
    \caption{Retrieval And Ranking Measurements}
    \label{table:evaluation}
\end{table*}

\subsubsection{Features}

\label{FeaturesSection}

Going into granular features and the exact empirical formulas used is out of the scope of this paper. The broad level features used while ranking are:
\begin{enumerate}
    \item \textbf{WordNet:} There are various features generated using WordNet \citep{Miller:1995:WLD:219717.219748} matching with questions and answers. This takes care of word-level semantics. For instance, if there is information about ``price of furniture" in a KB and the end-user asks about ``price of table", the user will likely get a relevant answer. The scores of these WordNet features are calculated as a function of:
    \begin{itemize}
        \item Distance of 2 words in the WordNet graph
        \item Distance of Lowest Common Hypernym from the root
        \item Knowledge-Base word importance (Local IDFs)
        \item Global word importance (Global IDFs)
    \end{itemize}
    This is the most important feature in our model as it has the highest relative feature gain.
    
    \item \textbf{CDSSM:} Convolutional Deep Structured Semantic Models \citep{huang2013learning} are used for sentence-level semantic matching. This is a dual encoder model that converts text strings (sentences, queries, predicates, entity mentions, etc) into their vector representations. These models are trained using millions of Bing Query Title Click-Through data. Using the \textit{source-model} for vectorizing user query and \textit{target-model} for vectorizing answer, we compute the cosine similarity between these two vectors, giving the relevance of answer corresponding to the query.
    
    \item \textbf{TF-IDF:} Though sentence-to-vector models are trained on huge datasets, they fail to effectively disambiguate KB specific data.
    This is where a standard TF-IDF~\cite{jones2000probabilistic} featurizer with local and global IDFs helps.
\end{enumerate}

\subsubsection{Contextual Features}
We extend the features for contextual ranking by modifying the candidate QAs and user query in these ways:
\begin{itemize}
    \item $Query_{modified}$ = Query + Previous Answer; For instance, if user query is ``yes" and the previous answer is ``do you want to know about XYZ", the current query becomes ``do you want to know about XYZ yes".
    \item Candidate QnA pairs are appended with its parent Questions and Answers; no contextual information is used from the user's query. For instance, if a candidate QnA has a question ``benefits" and its parent question was ``know about XYZ", the candidate QA's question is changed to ``know about XYZ benefits".
\end{itemize}
The features mentioned in Section~\ref{FeaturesSection} are calculated for the above combinations also. These features carry contextual information.
\subsubsection{Modeling and Training} \label{sec:model_info}
We use gradient-boosted decision trees as our ranking model to combine all the features. Early stopping ~\cite{zhang2005boosting} based on Generality-to-Progress ratio is used to decide the number of step trees and Tolerant Pruning \cite{tamon2000boosting} helps prevent overfitting. We follow incremental training if there is small changes in features or training data so that the score distribution is not changed drastically.

\subsection{Persona Based Chit-Chat}

We add support for bot-developers to directly enable handling chit-chat queries like ``hi", ``thank you", ``what's up" in their QnAMaker bots. In addition to chit-chat, we also give bot developers the flexibility to ground responses for such queries in a specific personality: professional, witty, friendly, caring, or enthusiastic. For example, the ``Humorous" personality can be used for a casual bot, whereas a ``Professional" personality is more suited in case of banking FAQs or task-completion bots. There is a list of 100+ predefined intents~\cite{DBLP:journals/corr/abs-1811-07600}. There is a curated list of queries for each of these intents, along with a separate query understanding layer for ranking these intents. The arbitration between chit-chat answers and user's knowledge base answers is handled by using a chat-domain classifier~\cite{Akasaki2017ChatDI}.

\subsection{Active Learning}
The majority of the KBs are created using existing FAQ pages or manuals but to improve the quality it requires effort from the developers. Active learning generates suggestions based on end-user feedback as well as ranker's implicit signals. For instance, if for a query, CDSSM feature was confident that one QnA should be ranked higher whereas wordnet feature thought other QnA should be ranked higher, active learning system will try to disambiguate it by showing this as a suggestion to the bot developer. To avoid showing similar suggestions to developers, DB-Scan clustering is done which optimizes the number of suggestions shown.


\section{Evaluation and Insights}
QnAMaker is not domain-specific and can be used for any type of data. To support this claim, we measure our system's performance for datasets across various domains. The evaluations are done by managed judges who understands the knowledge base and then judge user queries relevance to the QA pairs (binary labels). Each query-QA pair is judged by two judges. We filter out data for which judges do not agree on the label. Chit-chat in itself can be considered as a domain. Thus, we evaluate performance on given KB both with and without chit-chat data (last two rows in Table~\ref{table:evaluation}), as well as performance on just chit-chat data (2nd row in Table ~\ref{table:evaluation}). Hybrid of deep learning(CDSSM) and machine learning features give our ranking model low computation cost, high explainability and significant F1/AUC score.
Based on QnAMaker usage, we observed these trends:
\begin{itemize}
    \item Around 27\% of the knowledge bases created use pre-built persona-based chitchat, out of which, $\sim$4\% of the knowledge bases are created for chit-chat alone. The highest used personality is Professional which is used in 9\% knowledge bases.
    \item Around $\sim$25\% developers have enabled active learning suggestions. The acceptance to reject ratio for active learning suggestions is 0.31.
    \item 25.5\% of the knowledge bases use one URL as a source while creation. $\sim$41\% of the knowledge bases created use different sources like multiple URLs. 15.19\% of the knowledge bases use both URL and editorial content as sources. Rest use just editorial content.
\end{itemize}

\section{Demonstration}

\begin{figure}
    \centering
    \includegraphics[width=0.48\textwidth]{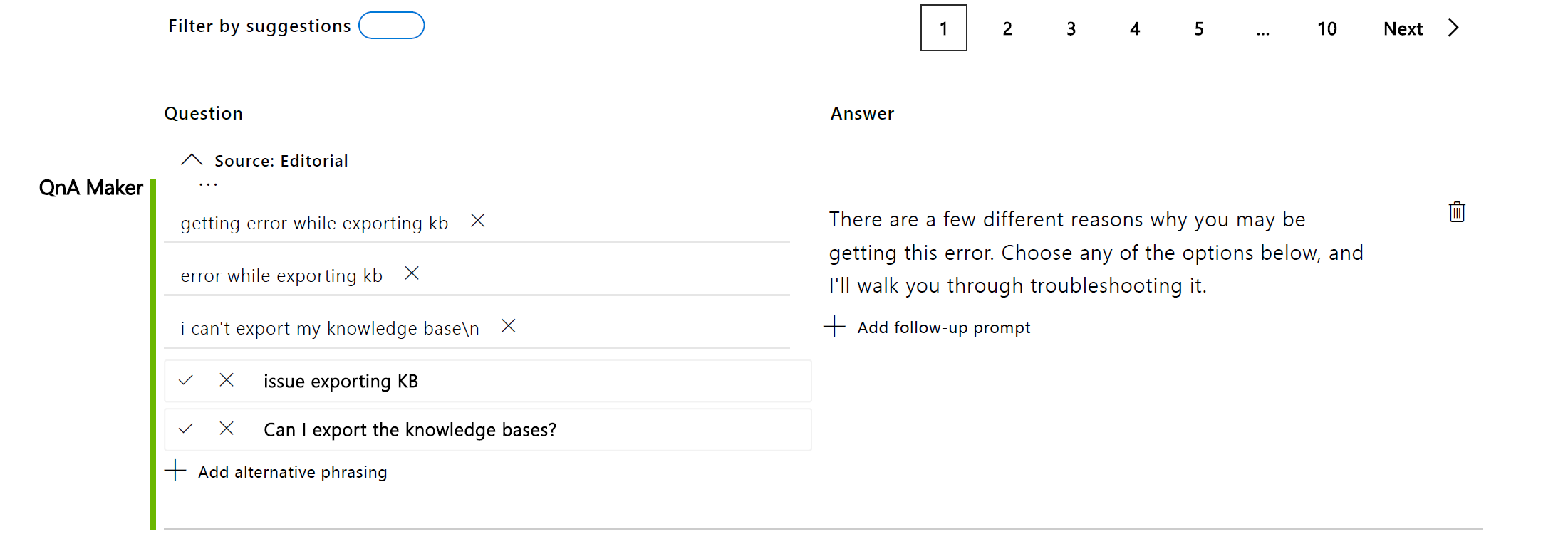}
    \caption{Active Learning Suggestions}
    \label{fig:SuggestionsKB}
\end{figure}

\begin{figure}
    \centering
    \includegraphics[width=0.48\textwidth]{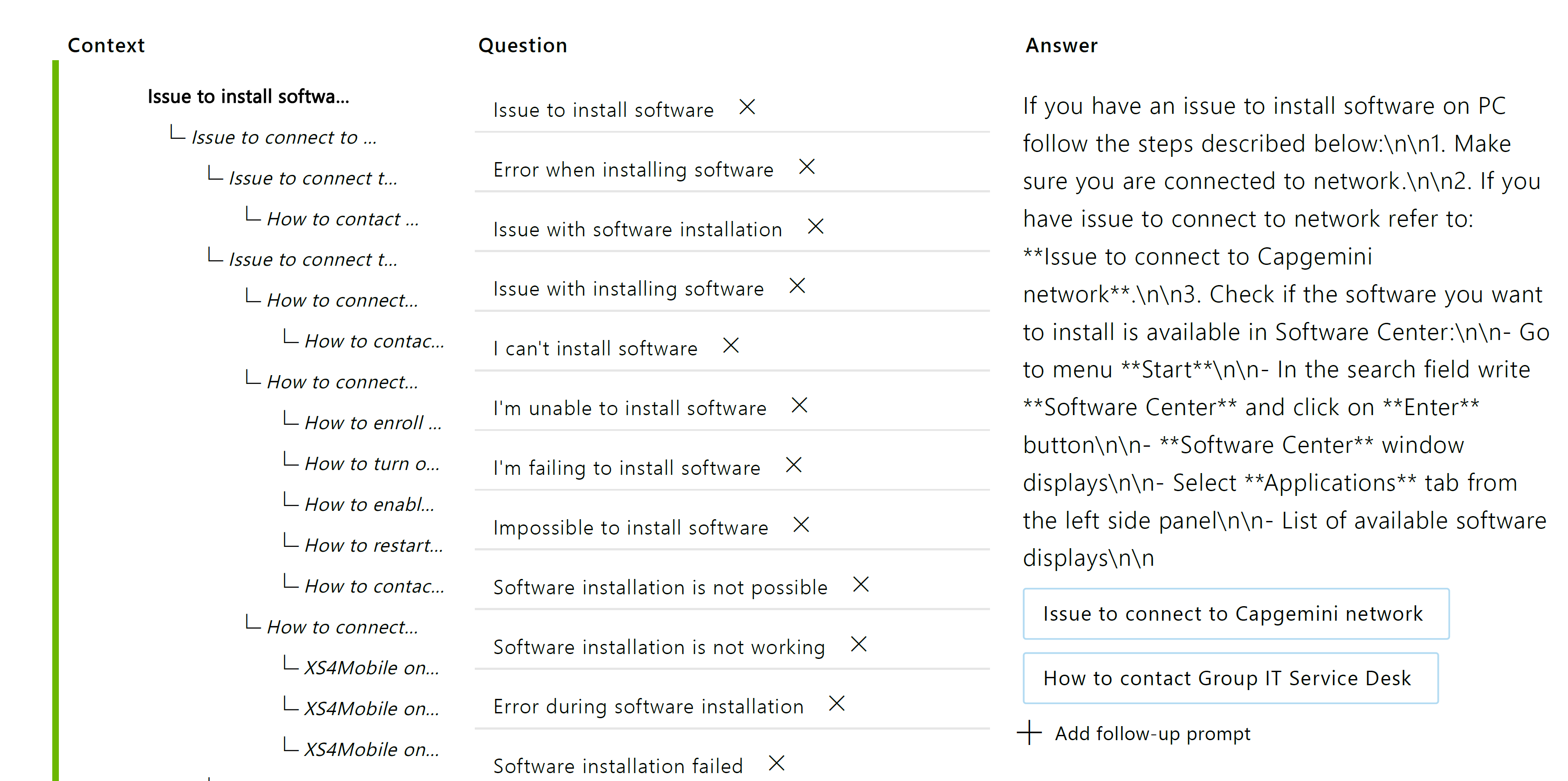}
    \caption{Multi-Turn Knowledge Base}
    \label{fig:MultiTurnKB}
\end{figure}

We demonstrate QnAMaker: a service to add a conversational layer over semi-structured user data. In addition to query-answering, we support novel features like personality-grounded chit-chat, active learning based on user-interaction feedback (Figure ~\ref{fig:SuggestionsKB}), and hierarchical extraction for multi-turn conversations (Figure ~\ref{fig:MultiTurnKB}). The goal of the demonstration will be to show how easy it is to create an intelligent bot using QnAMaker. All the demonstrations will be done on the production website \footnote{http://www.qnamaker.ai}
Demo Video can be seen here. \footnote{https://youtu.be/nBmBpsjuDOo}

\section{Future Work}The system currently doesn't highlight the answer span and does not generate answers taking the KB as grounding. We will be soon supporting Answer Span~\cite{NIPS2015_5945} and KB-grounded response generation~\cite{qin2019conversing} in QnAMaker. We are also working on user-defined personas for chit-chat (automatically learned from user-documents). We aim to enhance our extraction to be able to work for any unstructured document as well as images. We are also experimenting on improving our ranking system by using semantic vector-based search as our retrieval and transformer-based models for re-ranking.



\bibliographystyle{ACM-Reference-Format}
\bibliography{sample-base}


\begin{thebibliography}{11}


\ifx \showCODEN    \undefined \def \showCODEN     #1{\unskip}     \fi
\ifx \showDOI      \undefined \def \showDOI       #1{#1}\fi
\ifx \showISBNx    \undefined \def \showISBNx     #1{\unskip}     \fi
\ifx \showISBNxiii \undefined \def \showISBNxiii  #1{\unskip}     \fi
\ifx \showISSN     \undefined \def \showISSN      #1{\unskip}     \fi
\ifx \showLCCN     \undefined \def \showLCCN      #1{\unskip}     \fi
\ifx \shownote     \undefined \def \shownote      #1{#1}          \fi
\ifx \showarticletitle \undefined \def \showarticletitle #1{#1}   \fi
\ifx \showURL      \undefined \def \showURL       {\relax}        \fi
\providecommand\bibfield[2]{#2}
\providecommand\bibinfo[2]{#2}
\providecommand\natexlab[1]{#1}
\providecommand\showeprint[2][]{arXiv:#2}

\bibitem[\protect\citeauthoryear{Agrawal, Suri, and Menon}{Agrawal
  et~al\mbox{.}}{2018}]%
        {DBLP:journals/corr/abs-1811-07600}
\bibfield{author}{\bibinfo{person}{Parag Agrawal}, \bibinfo{person}{Anshuman
  Suri}, {and} \bibinfo{person}{Tulasi Menon}.}
  \bibinfo{year}{2018}\natexlab{}.
\newblock \showarticletitle{A Trustworthy, Responsible and Interpretable System
  to Handle Chit Chat in Conversational Bots}.
\newblock \bibinfo{journal}{\emph{CoRR}}  \bibinfo{volume}{abs/1811.07600}
  (\bibinfo{year}{2018}).
\newblock
\showeprint[arxiv]{1811.07600}
\urldef\tempurl%
\url{http://arxiv.org/abs/1811.07600}
\showURL{%
\tempurl}


\bibitem[\protect\citeauthoryear{Akasaki and Kaji}{Akasaki and Kaji}{2017}]%
        {Akasaki2017ChatDI}
\bibfield{author}{\bibinfo{person}{Satoshi Akasaki} {and}
  \bibinfo{person}{Nobuhiro Kaji}.} \bibinfo{year}{2017}\natexlab{}.
\newblock \showarticletitle{Chat Detection in an Intelligent Assistant:
  Combining Task-oriented and Non-task-oriented Spoken Dialogue Systems}. In
  \bibinfo{booktitle}{\emph{ACL}}.
\newblock


\bibitem[\protect\citeauthoryear{Cutting, Karger, Pedersen, and Tukey}{Cutting
  et~al\mbox{.}}{1992}]%
        {cutting1992scatter}
\bibfield{author}{\bibinfo{person}{Douglass~R Cutting},
  \bibinfo{person}{David~R Karger}, \bibinfo{person}{Jan~O Pedersen}, {and}
  \bibinfo{person}{John~W Tukey}.} \bibinfo{year}{1992}\natexlab{}.
\newblock \showarticletitle{Scatter/gather: A cluster-based approach to
  browsing large document collections}. In
  \bibinfo{booktitle}{\emph{Proceedings of the 15th annual international ACM
  SIGIR conference on Research and development in information retrieval}}. ACM,
  \bibinfo{pages}{318--329}.
\newblock


\bibitem[\protect\citeauthoryear{Ha, Haralick, and Phillips}{Ha
  et~al\mbox{.}}{1995}]%
        {ha1995recursive}
\bibfield{author}{\bibinfo{person}{Jaekyu Ha}, \bibinfo{person}{Robert~M
  Haralick}, {and} \bibinfo{person}{Ihsin~T Phillips}.}
  \bibinfo{year}{1995}\natexlab{}.
\newblock \showarticletitle{Recursive XY cut using bounding boxes of connected
  components}. In \bibinfo{booktitle}{\emph{Proceedings of 3rd International
  Conference on Document Analysis and Recognition}}, Vol.~\bibinfo{volume}{2}.
  IEEE, \bibinfo{pages}{952--955}.
\newblock


\bibitem[\protect\citeauthoryear{Hermann, Kocisky, Grefenstette, Espeholt, Kay,
  Suleyman, and Blunsom}{Hermann et~al\mbox{.}}{2015}]%
        {NIPS2015_5945}
\bibfield{author}{\bibinfo{person}{Karl~Moritz Hermann}, \bibinfo{person}{Tomas
  Kocisky}, \bibinfo{person}{Edward Grefenstette}, \bibinfo{person}{Lasse
  Espeholt}, \bibinfo{person}{Will Kay}, \bibinfo{person}{Mustafa Suleyman},
  {and} \bibinfo{person}{Phil Blunsom}.} \bibinfo{year}{2015}\natexlab{}.
\newblock \showarticletitle{Teaching Machines to Read and Comprehend}.
\newblock In \bibinfo{booktitle}{\emph{Advances in Neural Information
  Processing Systems 28}}, \bibfield{editor}{\bibinfo{person}{C.~Cortes},
  \bibinfo{person}{N.~D. Lawrence}, \bibinfo{person}{D.~D. Lee},
  \bibinfo{person}{M.~Sugiyama}, {and} \bibinfo{person}{R.~Garnett}} (Eds.).
  \bibinfo{publisher}{Curran Associates, Inc.}, \bibinfo{pages}{1693--1701}.
\newblock
\urldef\tempurl%
\url{http://papers.nips.cc/paper/5945-teaching-machines-to-read-and-comprehend.pdf}
\showURL{%
\tempurl}


\bibitem[\protect\citeauthoryear{Huang, He, Gao, Deng, Acero, and Heck}{Huang
  et~al\mbox{.}}{2013}]%
        {huang2013learning}
\bibfield{author}{\bibinfo{person}{Po-Sen Huang}, \bibinfo{person}{Xiaodong
  He}, \bibinfo{person}{Jianfeng Gao}, \bibinfo{person}{Li Deng},
  \bibinfo{person}{Alex Acero}, {and} \bibinfo{person}{Larry Heck}.}
  \bibinfo{year}{2013}\natexlab{}.
\newblock \showarticletitle{Learning deep structured semantic models for web
  search using clickthrough data}. In \bibinfo{booktitle}{\emph{Proceedings of
  the 22nd ACM international conference on Conference on information \&
  knowledge management}}. ACM, \bibinfo{pages}{2333--2338}.
\newblock


\bibitem[\protect\citeauthoryear{Jones, Walker, and Robertson}{Jones
  et~al\mbox{.}}{2000}]%
        {jones2000probabilistic}
\bibfield{author}{\bibinfo{person}{K~Sparck Jones}, \bibinfo{person}{Steve
  Walker}, {and} \bibinfo{person}{Stephen~E. Robertson}.}
  \bibinfo{year}{2000}\natexlab{}.
\newblock \showarticletitle{A probabilistic model of information retrieval:
  development and comparative experiments: Part 2}.
\newblock \bibinfo{journal}{\emph{Information processing \& management}}
  \bibinfo{volume}{36}, \bibinfo{number}{6} (\bibinfo{year}{2000}),
  \bibinfo{pages}{809--840}.
\newblock


\bibitem[\protect\citeauthoryear{Miller}{Miller}{1995}]%
        {Miller:1995:WLD:219717.219748}
\bibfield{author}{\bibinfo{person}{George~A. Miller}.}
  \bibinfo{year}{1995}\natexlab{}.
\newblock \showarticletitle{WordNet: A Lexical Database for English}.
\newblock \bibinfo{journal}{\emph{Commun. ACM}} \bibinfo{volume}{38},
  \bibinfo{number}{11} (\bibinfo{date}{Nov.} \bibinfo{year}{1995}),
  \bibinfo{pages}{39--41}.
\newblock
\showISSN{0001-0782}
\urldef\tempurl%
\url{https://doi.org/10.1145/219717.219748}
\showDOI{\tempurl}


\bibitem[\protect\citeauthoryear{Qin, Galley, Brockett, Liu, Gao, Dolan, Choi,
  and Gao}{Qin et~al\mbox{.}}{2019}]%
        {qin2019conversing}
\bibfield{author}{\bibinfo{person}{Lianhui Qin}, \bibinfo{person}{Michel
  Galley}, \bibinfo{person}{Chris Brockett}, \bibinfo{person}{Xiaodong Liu},
  \bibinfo{person}{Xiang Gao}, \bibinfo{person}{Bill Dolan},
  \bibinfo{person}{Yejin Choi}, {and} \bibinfo{person}{Jianfeng Gao}.}
  \bibinfo{year}{2019}\natexlab{}.
\newblock \showarticletitle{Conversing by Reading: Contentful Neural
  Conversation with On-demand Machine Reading}. In
  \bibinfo{booktitle}{\emph{Proc. of ACL}}.
\newblock
\urldef\tempurl%
\url{https://www.microsoft.com/en-us/research/publication/conversing-by-reading-contentful-neural-conversation-with-on-demand-machine-reading/}
\showURL{%
\tempurl}


\bibitem[\protect\citeauthoryear{Tamon and Xiang}{Tamon and Xiang}{2000}]%
        {tamon2000boosting}
\bibfield{author}{\bibinfo{person}{Christino Tamon} {and} \bibinfo{person}{Jie
  Xiang}.} \bibinfo{year}{2000}\natexlab{}.
\newblock \showarticletitle{On the boosting pruning problem}. In
  \bibinfo{booktitle}{\emph{European conference on machine learning}}.
  Springer, \bibinfo{pages}{404--412}.
\newblock


\bibitem[\protect\citeauthoryear{Zhang, Yu, et~al\mbox{.}}{Zhang
  et~al\mbox{.}}{2005}]%
        {zhang2005boosting}
\bibfield{author}{\bibinfo{person}{Tong Zhang}, \bibinfo{person}{Bin Yu},
  {et~al\mbox{.}}} \bibinfo{year}{2005}\natexlab{}.
\newblock \showarticletitle{Boosting with early stopping: Convergence and
  consistency}.
\newblock \bibinfo{journal}{\emph{The Annals of Statistics}}
  \bibinfo{volume}{33}, \bibinfo{number}{4} (\bibinfo{year}{2005}),
  \bibinfo{pages}{1538--1579}.
\newblock


\end{thebibliography}

\end{document}